\documentclass[8pt]{article}  \usepackage{times}
\usepackage{graphicx}

\usepackage{color}

\usepackage[round,numbers,sort&compress]{natbib} 

\usepackage{amsmath}
\usepackage{amssymb}
\usepackage{stackrel}
\usepackage{chemarrow}
\usepackage{graphicx}

\usepackage{booktabs}
\usepackage{dcolumn}
\usepackage{rotating}
\usepackage{multirow}

\newcommand{\be}{\begin{equation}}
\newcommand{\ee}{\end{equation}}

\let\baraccent=\= 
\renewcommand{\=}[1]{\stackrel{#1}{=}} 

\newcommand{\eref}[1]{eq.~(\ref{#1})}
\newcommand{\fref}[1]{fig.~\ref{#1}}

\usepackage[euler]{textgreek}

\begin{document}

\twocolumn[{\LARGE \textbf{Electrical properties of polar membranes\\*[0.2cm]}}
{\large Lars D. Mosgaard$^{\dagger}$, Karis A.Zecchi$^{\dagger}$, Thomas Heimburg$^{\ast}$\\*[0.1cm]
{\small Niels Bohr Institute, University of Copenhagen, Blegdamsvej 17, 2100 Copenhagen \O, Denmark}\\

{\normalsize ABSTRACT\hspace{0.5cm} Biological membranes are capacitors that can be charged by applying a field across the membrane. The charges on the capacitor exert a force on the membrane that leads to electrostriction, i.e. a thinning of the membrane. Since the force is quadratic in voltage, negative and positive voltage have an identical influence on the physics of symmetric membranes. However, this is not the case for a membrane with an asymmetry leading to a permanent electric polarization. Positive and negative voltages of identical magnitude lead to different properties. Such an asymmetry can originate from a lipid composition that is different on the two monolayers of the membrane, or from membrane curvature. The latter effect is called 'flexoelectricity'. As a consequence of permanent polarization, the membrane capacitor is discharged at a voltage different from zero. This leads to interesting electrical phenomena such as outward or inward rectification of membrane permeability.

Here, we introduce a generalized theoretical framework, that treats capacitance, polarization, flexoelectricity and piezoelectricity in the same language.
\\*[0.0cm] }}
]

\noindent\footnotesize {$^{\ast}$corresponding author, theimbu@nbi.dk. $^{\dagger}$LDM and KAZ contributed equally to this work.}\\


\normalsize
\section*{Introduction}
Many signaling  processes in biology involve electrical phenomena. These processes are related to the movement of ions and the orientation of polar molecules. Biological molecules typically contain charged groups that are at the origin of electrical fields and dipole moments. Furthermore, membranes and macromolecules are surrounded by electrolytes containing char{-}ged ions.  At physiological ionic strength, the Debye length of electrostatic interactions in the aqueous medium is about 1\;nm. It is caused by the shielding of charges by ions. However, in the hydrophobic cores of membranes and proteins, the dielectric constant is small, and no ions that could shield electrostatic interactions are present. Thus, the length scale of electrostatic interactions is significantly larger. Generally, under physiological conditions the range of the electric fields is similar to the size of biological macromolecules.  In this publication we will focus on the electrostatics of membranes that determines capacitance, polarization, piezoelectricity and flexoelectricity.

\begin{figure}[!b]
	\centering
		\includegraphics[width=  1 \linewidth]{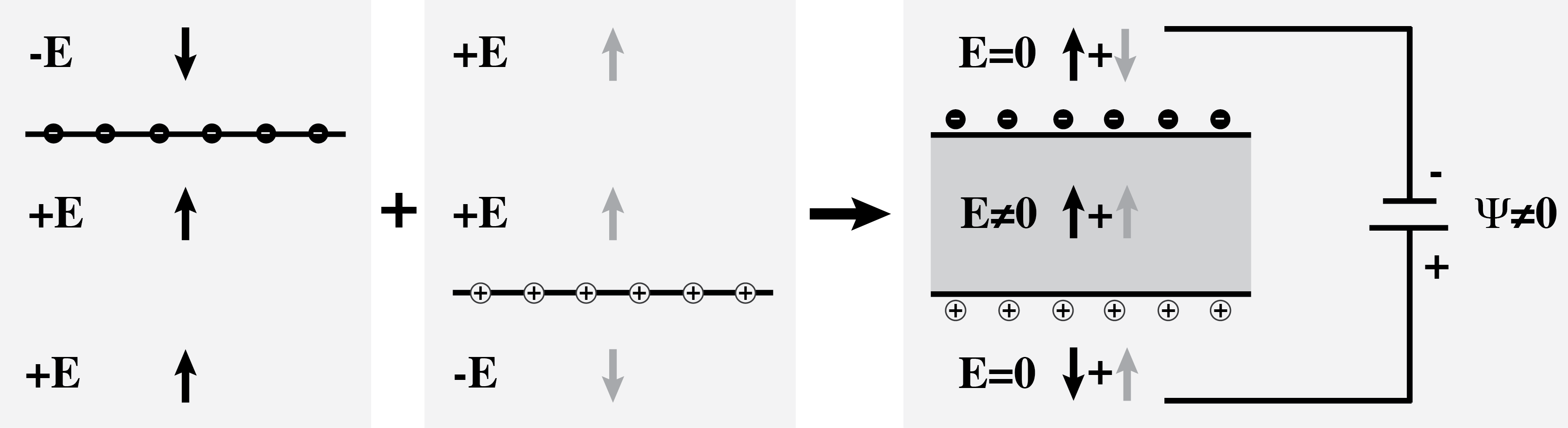}
	\parbox[c]{8 cm}{ \caption{Illustration of capacitive effects. The field inside a charged capacitor can be obtained by the superposition of the fields of a positively and a negatively charged plate at distance d. The charged capacitor displays an internal field different from zero, while the field is zero outside of the capacitor.
		\label{Figure0a}}}    
\end{figure}

There exist large concentration differences of ions across the membranes of biological cells. For instance, the concentration of potassium is about 400\;mM inside and only 20\;mM outside of a squid axon. If the membrane is selective for potassium, this results in a Nernst potential across the biological membrane. The combination of  the Nernst potentials of different ions yields a resting potential, which for biological cells is in the range of $\pm 100$ mV. The central core of a membrane is mostly made of hydrophobic non-conductive material. Thus, the biomembrane is considered a capacitor, e.g., in the Hodgkin-Huxley model for the nervous impulse \cite{Hodgkin1952b}. During the nerve pulse, currents are thought to flow across ion channel proteins that transiently charge or discharge the membrane capacitor. Within this model, the membrane is assumed to be a homogeneous planar capacitor with constant dimensions. The capacitance can be calculated from the relation 
\be
C_m=\varepsilon \cdot \frac{A}{d}
\label{cap1}
\ee
where $\varepsilon$ is the dielectric constant, $A$ is the membrane area and $d$ is the membrane thickness.

Let us assume that the membrane is surrounded by a conducting electrolyte solution. In the presence of an applied voltage, the charged capacitor consists of one plate with positive charges and one plate with negative charges at distance $d$. The field inside the capacitor can be determined using the superposition of the fields of the two plates (illustrated in Fig. \ref{Figure0a}). The field inside a charged capacitor is different from zero, while it is zero outside of the capacitor. If no field is applied, the capacitor is not charged.

\begin{figure}[!h]
	\centering
		\includegraphics[width= 1 \linewidth]{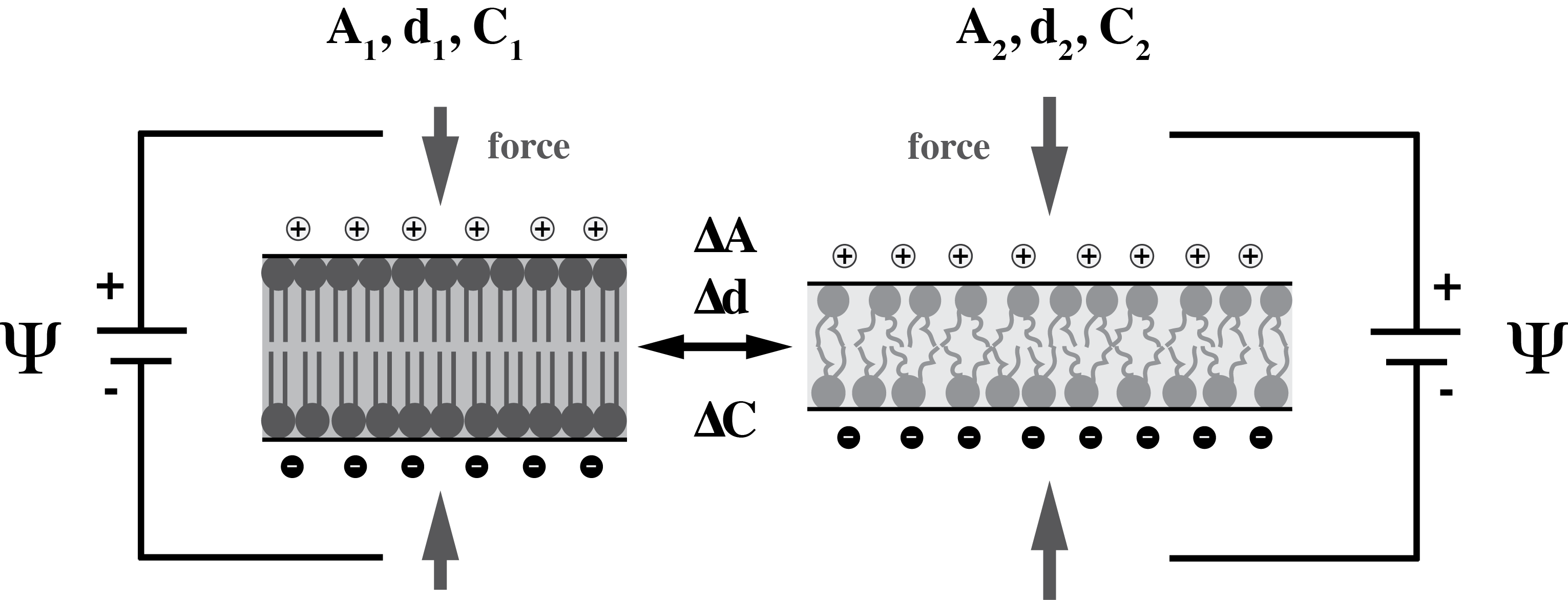}
	\parbox[c]{8 cm}{ \caption{Illustration of the electrostriction effect upon charging the membrane capacitor. The potential difference, $\Psi$, results in a force on the membrane that leads to a compression of the membrane to a state with larger area, $A$, and lower thickness, $d$.
	\label{Figure1a}}}    
\end{figure}

The charges on a capacitor generate mechanical forces on the two membrane layers \cite{Heimburg2012}. These forces can change the dimensions of the capacitor such that both the area and the thickness of the membrane change (see \fref{Figure1a}). As a result, the  capacitance is not generally a constant \cite{Heimburg2012}. The capacitance increases upon charging the membrane by an applied field because the membrane thickness decreases and the area increases. This effect is known as electrostriction. Close to phase transitions in the membrane  (in which the compressibility of the membrane is large \cite{Heimburg1998}), the membrane should be considered as a nonlinear capacitor. A small change in voltage can result in large changes in thickness and capacitance. The coupling between the membrane voltage and its dimensions renders the membrane piezoelectric, i.e., mechanical changes in the membrane can create a membrane potential and vice versa.\\

On average, about 80\% of the lipids are zwitterionic. Zwitterionic lipids possess permanent electrical dipole moments. Examples of such lipids are phosphatidylcholines and phosphatidylethanolamines. About 10\% of biological lipids carry a net negative charge, including phosphatidylinositol and phosphatidylserine. It is known that biomembranes often display asymmetric distributions of lipids such that charged lipids are mostly found in the inner leaflet of the bilayer \cite{Rothman1977}. Biomembranes also contain integral and peripheral proteins with asymmetric distribution (or orientation ) between inside and outside, which carry both positive and negative charges. Due to such compositional asymmetries, a spontaneous electrical dipole moment of the membrane can be generated in the absence of an externally applied field. 
\begin{figure}[!h]
	\centering
		\includegraphics[width= 1 \linewidth]{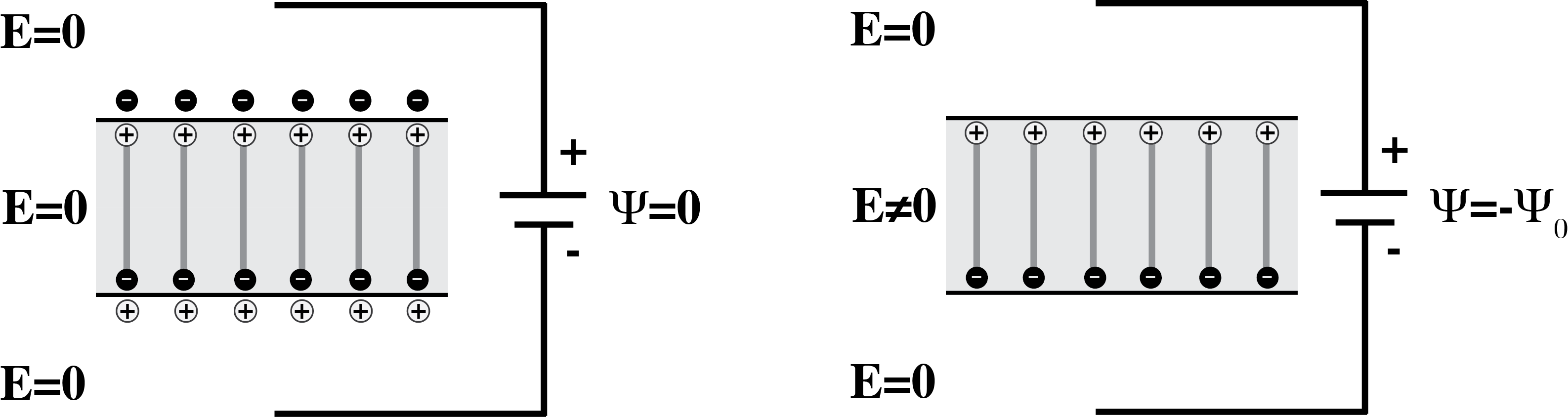}
	\parbox[c]{8 cm}{ \caption{Illustration of polarization by chemical asymmetry. Left: If the membrane contains permanent electrical dipoles, it is charged even if the applied potential is zero. Both, the fields inside and outside of the capacitor are zero. Right: In order to discharge the capacitor, a potential of $\Psi=-\Psi_0$ has to be applied.
		\label{Figure0b}}}    
\end{figure}
A redistribution or reorientation of polar molecules in an external field resembles the charging of a capacitor. If the membrane possesses a spontaneous polarization, the membrane capacitor in equilibrium can be charged even in the absence of an external field (illustrated in Fig. \ref{Figure0b}). In order to discharge this capacitor, a potential of $\Psi=-\Psi_0$ has to applied. We call $\Psi_0$ the spontaneous membrane potential, or the offset potential. In a theoretical treatment one has to be very careful to correctly account for both capacitive and polarization effects.

\begin{figure}[!t]
	\centering
		\includegraphics[width= 1 \linewidth]{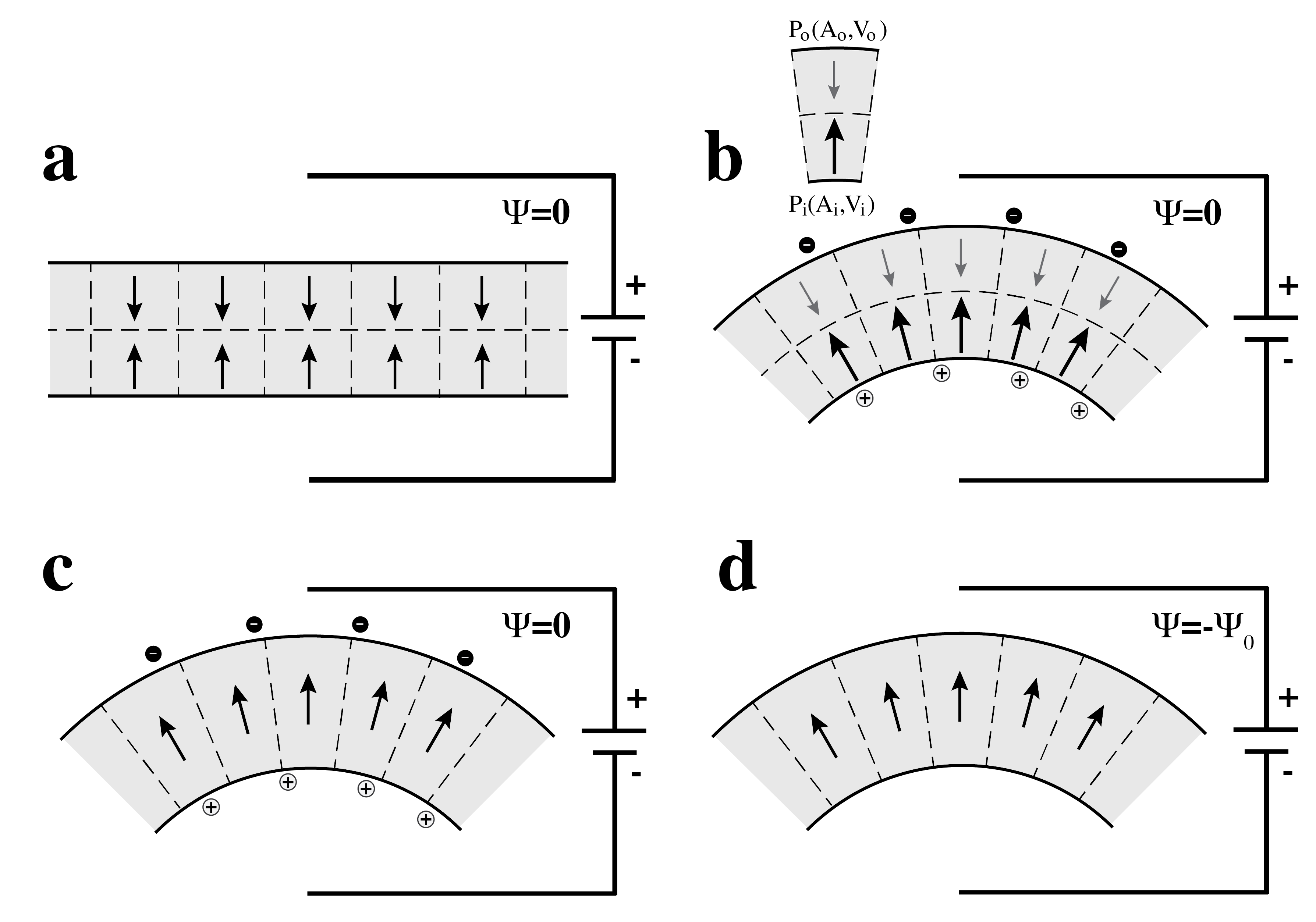}
	\parbox[c]{8 cm}{ \caption{Illustration of polarization by curvature. a. The two monolayers of the symmetric membrane display opposite polarization. b. upon bending (flexing) the membrane, the polarization in the two layers changes. c. effective polarization of the membrane. d: In order to discharge the capacitor, a potential of $\Psi=-\Psi_0$ has to be applied.
		\label{Figure0c}}}    
\end{figure}

The polarization effects described above rely on an asymmetric distribution of charges or dipoles on the two sides of a membrane.  Interestingly, even a chemically symmetric lipid membrane made of zwitterionic (uncharged) lipid may be polarized. The individual monolayers of zwitterionic lipids display trans-layer voltages on the order of 300 mV \cite{Vogel1988, Brockman1994}.  Any geometric deformation that breaks the symmetry between the two monolayers of a membrane results in a net polarization if these distortions alter the relative dipole orientation on the two layers. In particular, curvature induces different lateral pressure on the two sides of a membrane. Thereby, curvature can induce polarization in the absence of an applied field. This consideration was introduced by Meyer in 1969 \cite{Meyer1969} for liquid crystals. It was applied to curved lipid bilayers by Petrov in 1975 \cite{Petrov1975}. He called this effect 'flexoelectricity'. Upon bending (or flexing) the membrane, both area, $A$, and volume, $V$, of the opposing monolayers change in opposite directions. If the polarization is a function of area and volume the polarization of the outer monolayer is given by $P_o\equiv P_o(A_o, V_o)$ and that of the inner monolayer is given by $P_i\equiv P_i(A_i, V_i)$, respectively. Therefore, curvature can induce a net polarization across the membrane. This is illustrated in Fig. \ref{Figure0c}. This polarization is counteracted by opposing charges adsorbing to the membranes (Fig. \ref{Figure0c}b and \ref{Figure0c}c). In order to discharge the membrane, a potential $\Psi=-\Psi_0$ has to be applied. As in the case of chemical asymmetry, at zero applied field, the field inside the capacitor is zero. The cases of a chemically asymmetric planar membrane and a chemically symmetric curved membrane are conceptually similar. 


Charged capacitors, polarization, flexoelectricity and piezoelectricity all involve the spatial separation of charges. Thus, they all represent aspects of the same electrostatic phenomena. However, in the literature they are often treated as different things and they are described by a different language. In this communication we formulate a general thermodynamical description of the electrostatics of lipid membranes, which represents a generalization of a study on the capacitance of membranes previously published by our group \cite{Heimburg2012}. It will be used to generalize the effect of an externally applied electric field on the lipid melting transition. We will introduce the thermodynamics of a polarized lipid membrane in an electric field, which then results in a generalization of electrostriction effects on lipid membranes.
1
\section*{Theory}\label{Theory}
When the molecules of dielectric materials are placed in an external electric field, they orient themselves to the free energy. In capacitors, net macroscopic dipoles are induced in the dielectric medium and tend to counteract the applied field. As a response to an applied electric field, mechanical changes can be observed, e.g., in piezoelectric crystals. To deal with these effects, authors like Frank treated the electrostatic effects within a thermodynamical framework \cite{Frank1955}. He considered the electrical work performed on a fluid during any infinitesimal and reversible change, $d W_{el}=Ed(v D)$. This type of consideration leads to expressing the electric displacement, $D$, in a volume, $v$, as an extensive variable with the electric field, $E$, as its conjugated intensive variable. Vector notation has been dropped assuming planar geometry.

When we consider a membrane capacitor,  its hydrophobic core separates the two capacitor plates and acts both as a compressible and dielectric material. Choosing hydrostatic pressure ($p$), lateral pressure ($\pi$), temperature ($T$) and applied electric field ($E$) as intensive variables, we can write the differential of the Gibbs free energy as 
\be
\text{d} G = -S \text{d}T + v \text{d} p + A \text{d} \pi - (vD) \text{d} E + ...
\label{Gibbs}
\ee
where the conjugated extensive variables are $S$ (entropy), $v$ (volume), $A$ (area) and $vD$ (electric displacement). The electrical contribution to the free energy due to an applied electric field comes from the final term, which we will refer to as the electrical free energy, $G^{el}$. 
\\\\
The electric displacement is related to the total polarization, $P_{tot}$ by
\be
D = \varepsilon_0 E + P_{tot}.
\label{dis0}
\ee
where $\varepsilon_0$ is the vacuum permittivity. Most materials have zero polarization at zero electric field, and polarization is only induced by an external field. For a linear dielectric material  the induced polarization is $P_{\text{ind}}=\varepsilon_0 \chi_{el}E$, where $\chi_{el}$ is the electric susceptibility. We are interested in extending our considerations to a dielectric material which can display spontaneous polarization, $P_0$, in the absence of an applied field such that
\be
P_{tot} =\varepsilon_0 \chi_{el}E + P_{0}.
\ee
The spontaneous polarization, $P_0$, can originate from asymmetric lipid bilayers, e.g., from curvature (flexoelectricity) or from different composition of the two monolayers. The electric displacement takes the form
\be
D = \varepsilon (E + E_0),
\label{dis1}
\ee
where $\varepsilon$ is the dielectric constant, $\varepsilon=\varepsilon_0(1+\chi_{el})$ and $E_0\equiv P_0/\varepsilon$ is the electric field related to the spontaneous polarization, $P_0$, at $E=0$. 
\\\\
Using \eref{dis1}, we can determine the the electrical free energy:
\begin{eqnarray}
G^{el} &=& -\int_0^E (vD) \text{d} E'= - \varepsilon v \left( \frac{E^2}{2} + E_0 E \right) \nonumber\\
&=& -\frac{\varepsilon}{2} v \left(( E + E_0)^2 - E_0^2 \right) \;,
\end{eqnarray}
where we have assumed the volume of the lipid membrane to be constant. Assuming that the dielectric properties of the medium are homogeneous across a membrane with thickness $d$, we can define $E d=\Psi$ where $\Psi$ represents the applied electric potential difference. This leads to
\be
G^{el} =  - \frac{\varepsilon}{2} \frac{A}{d} \left( ( \Psi + \Psi_0)^2 - \Psi_0^2 \right) \;,
\label{G_el}
\ee
where $\Psi_0$ is the offset potential related to $E_0$ ($E_0 d=\Psi_0$).
The pre-factor contains the capacitance of a planar capacitor ($C_m = \varepsilon A/d$). Thus, the electric free energy is given by
\be
\label{G_el_1}
G^{el} =  - \frac{1}{2} C_m \left( ( \Psi + \Psi_0)^2 - \Psi_0^2 \right). 
\ee
At $\Psi=0$ the electrical contribution to the free energy is zero.

\subsection*{Electrostriction}
The charges on a capacitor attract each other. These attractive forces can change the dimensions of the membrane and thereby change the capacitance. If $\Psi_0=0$, the electric contribution to the free energy according to eq. (\ref{G_el_1}) is $G^{el} =  - \frac{1}{2} C_m  \Psi^2$. For $A\approx $ const. and $\Psi= $ const., the force $\mathcal{F}$ acting on the layers is
\be
\mathcal{F}=\frac{\partial G^{el}}{\partial d} = -\frac{1}{2} \left(\frac{\partial C_m}{\partial d}\right) \Psi^2=\frac{1}{2}\frac{C_m\Psi^2}{d} \;.
\ee
This is the force acting on a planar capacitor given in the literature (e.g., \cite{Heimburg2012}). If there exists a constant offset potential $\Psi_0$, we find instead (eq. (\ref{G_el_1}))
\be
\mathcal{F}=\frac{1}{2}\frac{C_m}{d}\left( ( \Psi + \Psi_0)^2 - \Psi_0^2 \right) \;.
\ee
\begin{figure}[!b]
	\centering
		\includegraphics[width= 1 \linewidth]{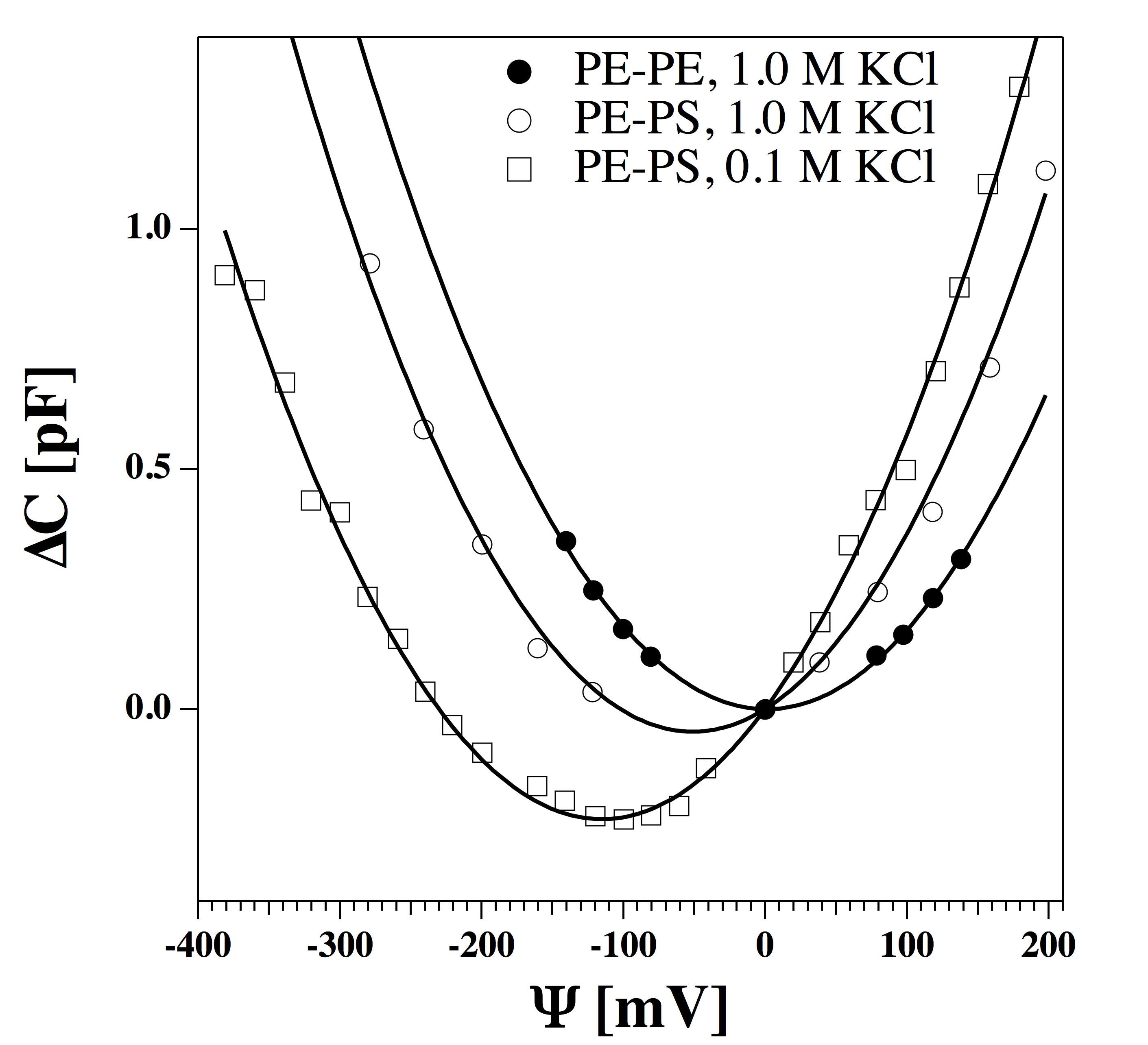}
	\parbox[c]{8cm}{ \caption{The change in capacitance as a function of potential in a black lipid membrane. Solid circles: Symmetric membrane in 1 M KCl. Both monolayers are made from zwitterionic bacterial phosphatidyl ethanolamine (PE).  Open circles: Asymmetric membrane in 1 M KCl. One monolayer is consists of bacterial PE, while the other monolayer consist of the charged bovine brain phosphatidylserine (PS). Open squares: Same as open circle, but with smaller salt concentration (0.1 M KCl).  The absolute capacitance, $C_{m,0}$ at $\Psi= 0$ V is approximately $300$ pF. Raw data adapted from \cite{Alvarez1978}.
	\label{Figure1d}}}    
\end{figure}
Thus, one expects that the force on a membrane is a quadratic function of voltage which displays an offset voltage when the membrane is polarized. This force can reduce the membrane thickness and thereby increase the capacitance of a membrane. Note, however, that for $( \Psi + \Psi_0)^2 - \Psi_0^2<0$, the force $\mathcal{F}$ is negative. As a consequence, capacitance will be decreased.

Let us assume a membrane with constant area and small thickness change, $\Delta d << d$. Then the change in capacitance, $\Delta C_m$, caused by a change of thickness, $\Delta d$, is given by
\be
\Delta C_m=-\varepsilon \frac{A}{d^2}\Delta d
\ee
Thus, the change in capacitance is proportional to the change in thickness. If the thickness is a linear function of the force ($\mathcal{F}\propto \Delta d$), one finds that the capacitance is proportional to the force $\mathcal{F}$. Therefore, it is a quadratic function of voltage with an offset of $\Psi_0$,

\be
\Delta C_m\propto \left( ( \Psi + \Psi_0)^2 - \Psi_0^2 \right) \;.
\label{capF}
\ee
The magnitude of the change in capacitance depends on the elastic constants of the membrane.

Relation (\ref{capF}) was studied by various authors. Using black lipid membranes, Alvarez and Latorre \cite{Alvarez1978} found a quadratic dependence of the capacitance on voltage (Fig. \ref{Figure1d}). In a symmetric membrane made of the zwitterionic (uncharged) lipid phosphatidylethanolamine (PE), the offset potential $\Psi_0$ in a 1 M KCl buffer was found to be zero. In an asymmetric membrane with one monolayer made of PE and the other made of the charged lipid phosphatidylserine (PS), a polarization is induced. In a 1 M KCl buffer, the offset potential is $\Psi_0= 47$ mV, while it is $\Psi_0=116$ mV in a 0.1 M KCl buffer. It is obvious from Fig. \ref{Figure1d} that within experimental error the shape of the capacitance profile is unaffected by the nature of the membrane. Only the offset potential is influenced by composition and ionic strength. This suggests that the offset potential has an ionic strength dependence. In this publication,  we do not explore the theoretical background of this experimental fact.

In a range of $\pm 300 mV$ around the minimum capacitance, the change in capacitance, $\Delta C_m$, is of the order of $< 1.5$ pF, while the absolute capacitance, $C_{m,0}$, at $\Psi=0$ is approximately 300 pF \cite{Alvarez1978}. Thus, the change in capacitance caused by voltage is very small compared to the absolute magnitude of the capacitance.

\subsection*{Influence of the potential on the capacitance close to a melting transition}
As discussed above, the influence of voltage on the capacitance is small in the gel and in the fluid phase because membranes are not very compressible in their pure phases. However, close to the phase transition between gel and fluid, membranes become very compressible. In this transition, the thickness of the membrane, $d$, decreases by about 16\% and the area, $A$, increases by about 24\% \cite{Heimburg1998} for the lipid dipalmitoyl phosphatidylcholine (DPPC). Therefore, the capacitance of the fluid membrane is about 1.5 time higher than the capacitance of the gel phase \cite{Heimburg2012}. According to eq.\;(\ref{G_el_1}), the Gibbs free energy difference caused by an external electric field can be written as
\be
\Delta G^{el} = G_{fluid}^{el} - G_{gel}^{el} = - \frac{\Delta C_m}{2} \left( ( \Psi + \Psi_0)^2 - \Psi_0^2 \right) \,,
\label{el-striction}
\ee
where $\Delta C_m $ is the difference between the capacitance of gel and fluid phase. Here, we assumed that both the offset potential $\Psi_0$ and the dielectric constant $\varepsilon$ do not change with the state. We have confirmed the latter in experiments on the dielectric constant in the melting transition of oleic acid using a parallel plate capacitor (data not shown). We found that the changes of the dielectric constant caused by the melting of oleic acid ($T_m\approx 17^\circ$C) are very small.
\begin{figure}[b!]
	\centering
		\includegraphics[width= 1 \linewidth]{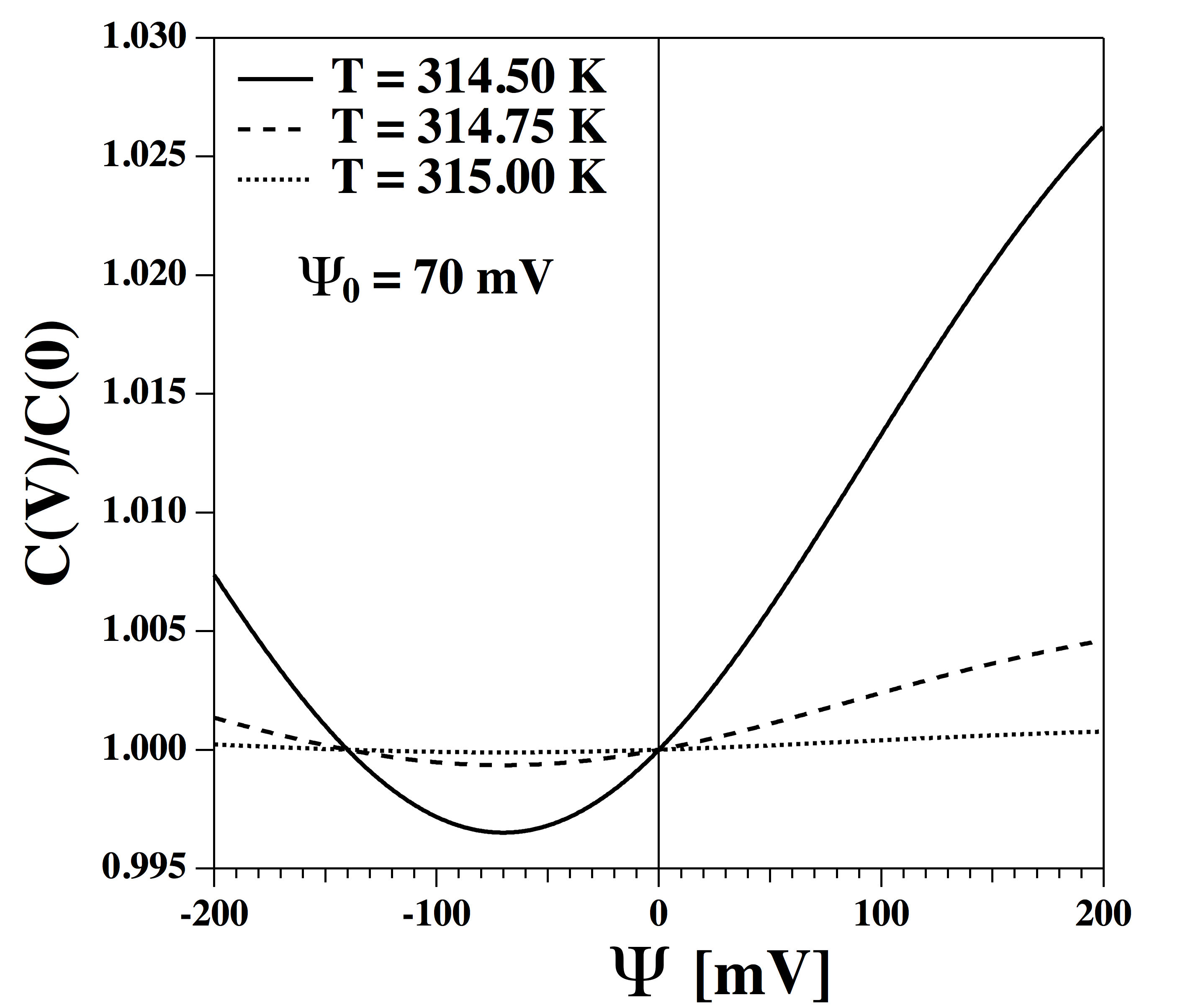}
	\parbox[c]{8cm}{ \caption{The relative change in capacitance of a lipid membrane as a function of applied voltage at three different temperatures above the melting temperature. Parameters are for LUV of DPPC, where $\Delta C \approx 656$ \; J/(mol$\cdot$V$^2$) and $\Psi_0$ was chosen to be $70$\;mV. 
	\label{Figure3}}}    
\end{figure}

It has been shown experimentally that in the vicinity of the lipid melting transition changes of various extensive variables are proportionally related \cite{Heimburg1998, Ebel2001, Schrader2002}. For instance, changes in enthalpy are proportional to changes in area, in volume and we assume that a similar relation holds for changes in thickness. Further, close to transitions the elastic constants are closely related to the heat capacity. For instance, the temperature-dependent change of the isothermal compressibility is proportional to heat capacity changes. Thus, membranes are more compressible close to transitions, and it is to be expected that the effect of potential changes on membrane capacitance is enhanced. This will be calculated in the following.
 
We assume that the lipid melting transition is described by a two-state transition governed by a van't Hoff law, so that the equilibrium constant between the gel and the fluid state of the membrane can be written as \cite{Heimburg2012, Graesboll2014}
\be
K(T, \Psi)=\exp \left(-n \frac{\Delta G}{RT} \right)
\label{freeEnergy1}
\ee 
where $n$ is the cooperative unit size which describes the number of lipids that change state cooperatively (for LUVs of DPPC we used $n=170$ \cite{Blicher2013}). The free energy difference between gel and fluid membranes is given by 
\be
\Delta G=(\Delta H_0-T\Delta S_0)+\Delta G^{el}\;,
\ee
where $\Delta H_0=35$\;kJ/mol and $\Delta S_0=111.4$\;J/mol K (for DPPC). From the equilibrium constant we can calculate the fluid fraction, the average fraction of the lipids that are in the fluid state,
\be
f_f(T, \Psi) = \frac{K(T, \Psi)}{1+K(T, \Psi)}.
\ee
For DPPC LUV, the thickness in the gel and fluid state is given by $d_{g} = 4.79 \; $nm  and $d_{f} = 3.92 \;$ nm , respectively. The area per lipid is $A_{g} = 0.474 \; $nm$^2 $ and $A_{f} =  0.629 \; $nm$^2 $ \cite{Heimburg1998}. We assume a dielectric constant of $\varepsilon = 4 \cdot \varepsilon_0$ independent of the state of the membrane. The area is described by $A(T, \Psi)=A_g+f_f\cdot \Delta A$, and the membrane thickness by $d(T, \Psi)=d_g-f_f\cdot \Delta d$, respectively. The temperature and voltage-dependent capacitance, $C=\varepsilon A(T, \Psi)/d(T, \Psi)$ is shown in \fref{Figure3}. For small variations in the potential, the change in capacitance is a quadratic function of voltage. For large potentials, one finds the capacitance of the fluid phase which is assumed being constant. One can recognize that the sensitivity of the capacitance to voltage changes close to the transition is much larger than that of the pure phases (Fig. \ref{Figure1d}). It is also a sensitive function of the temperature. Fig. \ref{Figure3} shows $C_m (\Psi)$ for three different temperatures above the melting temperature of DPPC at 314.15 $^\circ$C. At $T=314.5$\;K, the change in capacitance at $\Psi-\Psi_0=300$ mV is approximately 3\% compared to the about $0.5$\% experimentally measured in the absence of a transition (Fig. \ref{Figure1d}). Due to the presence of a melting transition, the curve profile in Fig. \ref{Figure3} is only a quadratic function of potential close to $\Psi=-\Psi_0$.

\begin{figure}[!h]
	\centering
		\includegraphics[width= 1 \linewidth]{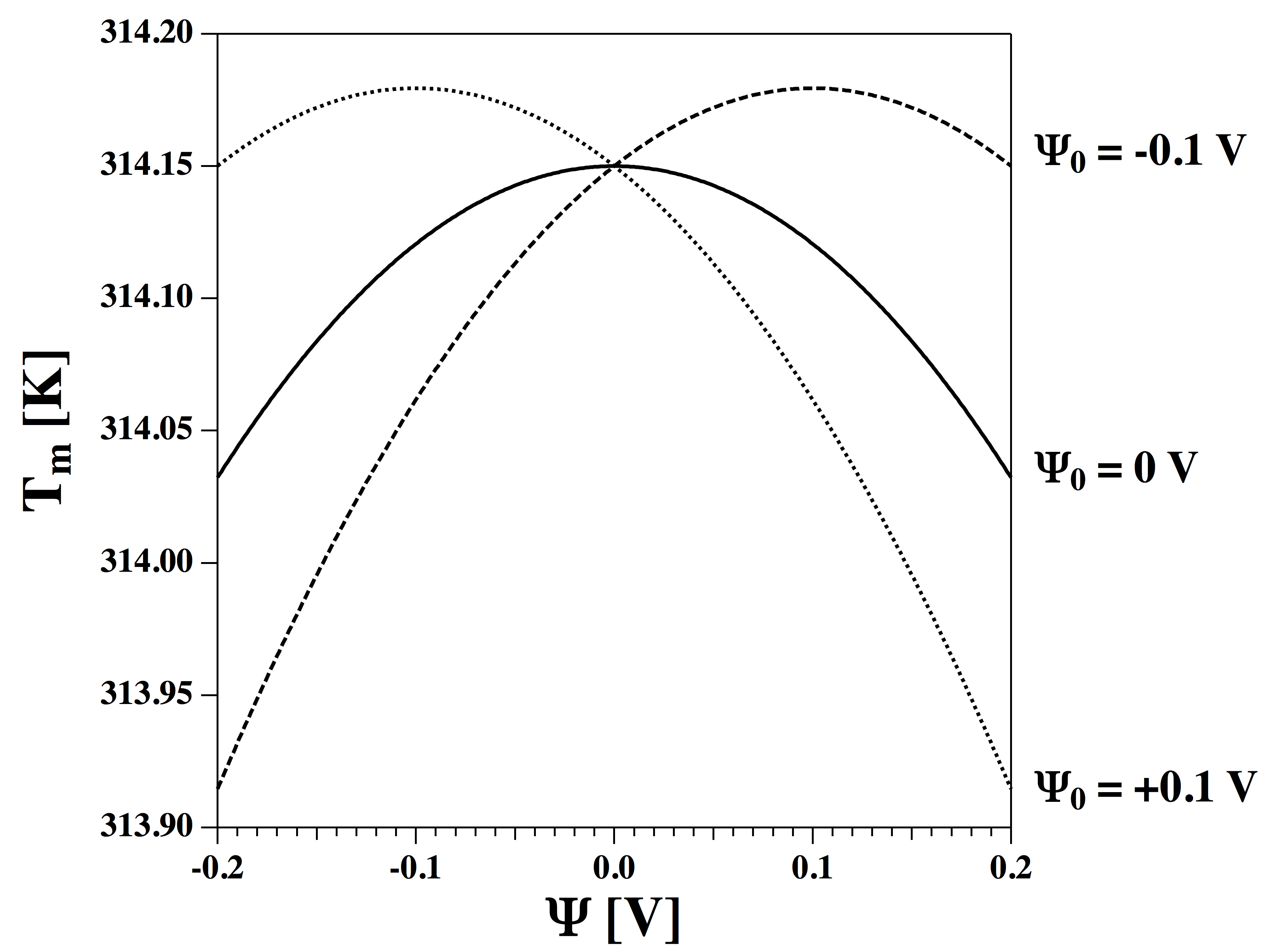}
	\parbox[c]{8cm}{ \caption{The lipid melting temperature as a function of applied potential with three different offset potentials, $\Psi_0$=0.1 V, $\Psi_0$=0V, and $\Psi_0$=-0.1 V. The parameters are taken from LUV of DPPC, where $\Delta C \approx 656 \; J/(mol\cdot V^2)$ for $\varepsilon = 4\cdot \varepsilon_0$. 
	\label{Figure1}}}    
\end{figure}
\subsubsection*{The dependence of the melting temperature on the applied potential}
The total free energy difference between gel and fluid phase, $\Delta G$, consists of an enthalpic and an entropic contribution,
\be
\Delta G = \Delta H_0 - T\Delta S_0+ \Delta G^{el}\,,
\label{DeltaG}
\ee
At the melting temperature, $T_m$, the Gibbs free energy difference $\Delta G$ is zero, so that
\begin{eqnarray}
T_m&=&T_{m,0}\left(1+\frac{\Delta G^{el}}{\Delta S_0}\right)  \\
&=&T_{m,0}\left(1-\frac{1}{2}\frac{\Delta C_m}{\Delta S_0} \left( ( \Psi + \Psi_0)^2 - \Psi_0^2 \right) \right)\,,\nonumber
\label{Tm}
\end{eqnarray}
where $T_{m,0}=\Delta H_0/\Delta S_0$ is the melting temperature in the absence of an external field (for DPPC: $\Delta H_0=35$ kJ/mol, $T_{m,0}=314.15$ K and $\Delta S_0=111.4$ J/mol$\cdot$ K \;\cite{Heimburg1998}). This result describes the effect of electrostriction on the lipid melting transition in the presence of spontaneous polarization. It is a generalization of the electrostriction effect described by Heimburg \cite{Heimburg2012} who treated this phenomenon in the absence of polarization effects. 
Fig. \ref{Figure1} shows the dependence of $T_m$ on an applied voltage for three different offset potentials, $\Psi_0$. It can be seen that in the presence of an applied filed, the spontaneous polarization and its sign influences that melting temperature.\\




\subsubsection*{Generalization for $\Psi_0 \ne $ const}

The orientation of lipid dipoles can change upon lipid melting. It seems to be obvious from lipid monolayer experiments that the polarization of liquid expanded and solid condensed layers is different. We assume the same to be true for bilayers.
Let us assume that the net offset potentials originating from membrane polarization in the gel and the fluid phase are given by $\Psi_0^g$ and $\Psi_0^f$, respectively. 
\begin{figure}[!h]
	\centering
		\includegraphics[width= 1 \linewidth]{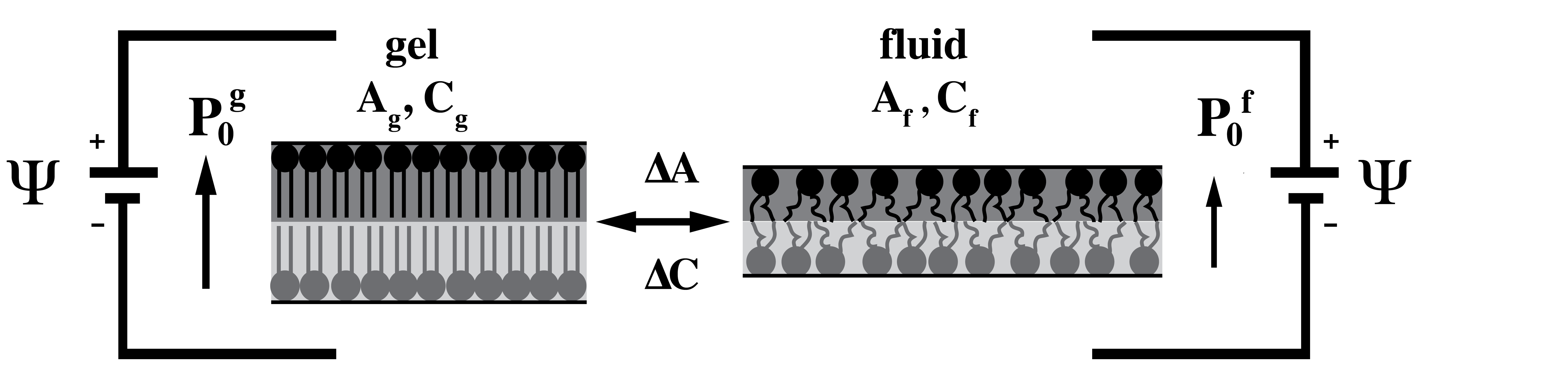}
	\parbox[c]{8cm}{ \caption{A polar membrane with different lipid composition on the top and bottom monolayer undergoing a melting transition from gel to fluid. The gel state possesses an area $A_g$ and a capacitance $C_g$, while the fluid state displays $A_f$ and $C_f$. The net offset potentials caused by membrane polarization are $\Psi_0^g$ and $\Psi_0^f$, respectively. The differences in area and capacitance between the two states are given by $\Delta A$ and $\Delta C$.
	\label{Figure2}}}    
\end{figure}
The free energy is now given by
\begin{eqnarray}
\Delta G^{el} &=& - \frac{C_f}{2} \left( ( \Psi + \Psi_{0,f})^2 - \Psi_{0,f}^2 \right) \nonumber\\
&&- \frac{C_g}{2} \left( ( \Psi + \Psi_{0,g})^2 - \Psi_{0,g}^2 \right) \nonumber\\
&=&-\frac{\Delta C}{2}((\Psi+\Psi_{0,g})^2-\Psi_{0,g}^2)\nonumber\\
&&-C_f\Psi(\Psi_{0,f}-\Psi_{0,g})
\label{piezo2}
\end{eqnarray}
This can be inserted in eq. (\ref{DeltaG}) to obtain the change in melting temperature due to an applied field.

\subsection*{The dielectric susceptibility}\label{dielectric_susceptibility}
In \cite{Heimburg2012} we defined a capacitive susceptibility, $\hat{C} =(\partial q/\partial V)=C+V(\partial C/\partial V)$. This susceptibility has a maximum at the melting temperature, which is a consequence of the fact that the capacitance of gel and fluid lipid phases differ. By analogy, we now introduce a dielectric susceptibility, $\hat{\varepsilon}=(\partial D/\partial E)$, which is given by:
\be
\hat{\varepsilon}\equiv\left(\frac{\partial D}{\partial E}\right)=\left(\frac{\partial(\varepsilon E+P_0)}{\partial E}\right)=\varepsilon+E\left(\frac{\partial \varepsilon}{\partial E}\right)+\left(\frac{\partial P_0}{\partial E}\right)
\label{suscept1}
\ee
Thermodynamic susceptibilities are linked to fluctuation relations. For instance, in \cite{Heimburg2012} we showed that the capacitive susceptibility is given by $\hat{C}=(\left<q^2\right>-\left<q\right>^2)/kT$, i.e., it is proportional to the fluctuations in charge. This fluctuation relation is valid as long as the distribution of states is described by Boltzmann statistics and the area and thickness are kept constant. Analogously, for constant volume, $v$, the dielectric susceptibility, $\hat{\varepsilon}$, is given by
\be
\hat{\varepsilon}=v\frac{\left<D^2\right>-\left<D\right>^2}{kT} \;.
\label{suscept2}
\ee
Since this is a positive definite form, $\hat{\varepsilon}$ is always larger than zero. The mean displacement, $\left<D\right>$, always increases with an increase in the electric field, $E$. If either $\varepsilon$ or the permanent polarization $P_0$ are different in the gel and the fluid state of a membrane, one can induce a transition. In this transition, the dielectric susceptibility displays an extremum. 


\section*{Capacitive susceptibility, piezoelectricity and flexoelectricity}\label{piezoelectricity}
As mentioned above, the polarization of a membrane can chan{-}ge by compressing, stretching or bending the membrane. The corresponding electrostatic phenomena are called electrostriction, piezoelectricity and flexoelectricity. In the past, some simple relations were derived by A. G. Petrov \cite{Petrov1999}. For instance, piezoelectricity was described as the area-dependence of polarization. Correspondingly, flexoelectricity was described as the curvature-dependence of the polarization assuming that polarization is zero in the planar state of the membrane. However, upon changing the membrane area, its capacitance also changes. Thus, in the presence of a field not only the polarization but also the charge on the capacitor can change. In the case of membrane curvature, the polarization may be different from zero in the planar state. Further, if there exists an applied potential, the capacitance of the membrane plays a role. In the following, we derive general equations for electrostriction, piezoelectricity and flexoelectricity. We will find that some relations previously derived by Petrov are special cases of our more general description. 

\subsection*{The charge on a capacitor}\label{ChargeCapacitor}

The dependence of the charge on a capacitor on potential, $\Psi$, surface area, $A$, and curvature, $c$ is given by
\be
d q=\left(\frac{\partial q}{\partial \Psi}\right)_{A,c} d\Psi + \left(\frac{\partial q}{\partial A}\right)_{\Psi,c} dA + \left(\frac{\partial q}{\partial c}\right)_{\Psi,A} dc\;.
\ee
Here, we assume that $\Psi$, $A$ and $c$ are variables that can be controlled in the experiment. The charge on a capacitor is given by
\be
q=A\cdot D = A(\varepsilon E +P_0) =\varepsilon \frac{A}{d} (\Psi+\Psi_0) = C_m (\Psi+\Psi_0)\;.
\ee
Thus, the change of the charge on a capacitor as a function of potential, lateral pressure, and curvature is given by:

\begin{eqnarray}
&&\left[(\Psi+\Psi_0)\left(\frac{\partial C_m}{\partial \Psi}\right)_{A,c} +C_m+C_m\left(\frac{\partial \Psi_0}{\partial \Psi}\right)_{A,c}\right] d\Psi \nonumber\\
dq&=&+ \left[(\Psi+\Psi_0)\left(\frac{\partial C_m}{\partial A}\right)_{\Psi,c} +C_m\left(\frac{\partial \Psi_0}{\partial A}\right)_{\Psi,c}\right] dA \\
&&+ \left[(\Psi+\Psi_0)\left(\frac{\partial C_m}{\partial c}\right)_{\Psi,A} +C_m\left(\frac{\partial \Psi_0}{\partial c}\right)_{\Psi,A}\right] d c \nonumber
\label{piezo_n0}
\end{eqnarray}
or in abbreviated form as
\begin{eqnarray}
&& \left[(\Psi+\Psi_0)\alpha_{A,c}+C_m+C_m\beta_{A,c}\right] d\Psi \nonumber\\
dq&\equiv&+ \left[(\Psi+\Psi_0)\alpha_{\Psi,c} +C_m\beta_{\Psi,c}\right] dA\qquad \;.\\
&&+ \left[(\Psi+\Psi_0)\alpha_{\Psi,A} +C_m\beta_{\Psi,A}\right] d c \nonumber
\label{piezo_n01}
\end{eqnarray}

The first term describes the change of charge on a capacitor allowing for the possibility that both capacitance and polarization can depend on voltage. The second term describes piezoelectricity, i.e., the change of charge by changing area, taking into account the area dependence of both capacitance and polarization.  The last term describes flexoelectricity, which relates to the change of charge caused by changes in curvature. Here, both dependence of capacitance and polarization on curvature are considered.

One could write similar equations, if the lateral pressure, $\pi$, were controlled instead of the area, $A$. 

\subsection*{Capacitive susceptibility}\label{Capacitance_general}
The capacitive susceptibility is given by $\hat{C}_m=\partial q/\partial \Psi$. It was discussed in detail in \cite{Heimburg2012}. In contrast to the capacitance, it can have a maximum in a melting transition. If lateral pressure and curvature are constant, we find from eq. (\ref{piezo_n01}) that
\be
\hat{C}_m=\left(\frac{\partial q}{\partial \Psi}\right)_{A, c}=(\Psi+\Psi_0)\alpha_{A,c}+C_m+C_m\beta_{A,c}
\label{cap_susc1}
\ee
If the spontaneous polarization is zero at all voltages, this reduces to
\be
\hat{C}_m=C_m+\Psi\left(\frac{\partial C_m}{\partial \Psi}\right)_{A,c} 
\ee
which is the relation given by Heimburg (2012). 

\subsection*{Piezoelectricity}\label{piezoelectricity0}

Let us assume that in eq. (\ref{piezo_n0}) $\Psi$ and $c$ are constant. We then obtain
\be
dq= \left[(\Psi+\Psi_0)\alpha_{\Psi,c} +C_m\beta_{\Psi,c}\right]dA\;.
\label{piezo_n1}
\ee
This effect is the 'piezoelectric effect'. It corresponds to the charging of a capacitor by changing the surface area of the membrane. At $\Psi=0$, we obtain for a small change in area, $\Delta A$,
\be
\Delta q\approx\left(\Psi_0\alpha_{\Psi,c}+C_m\beta_{\Psi,c}\right)\Delta A \;.
\ee
If $\Psi_0 (\Delta A=0)$ is zero, the capacitor is uncharged for $\Psi=0$. Then the charge on the capacitor after a change in area of $\Delta A$ is given by
\be
q (\Delta A)= C_m\beta_{\Psi,c}\Delta A \qquad\mbox{or}\qquad \Psi_0(\Delta A)=\beta_{\Psi,c}\Delta A \;.
\ee
A similar relation was given by Petrov and Usherwood \cite{Petrov1994}.\\


\noindent\textbf{Inverse piezoelectric effect:} The elastic free energy density of membrane compression is given by $g=\frac{1}{2}K_T^A (\Delta A/A_0)^2$, where $K_T^A$ is the lateral compression modulus and $A_0$ is the equilibrium area prior to compression. In the presence of an applied potential, the free energy  is given by
\be
g=\frac{1}{2}K_T^A \left(\frac{\Delta A}{A_0}\right)^2- \frac{1}{2} \frac{C_m}{A_0} \left( ( \Psi + \Psi_0)^2 - \Psi_0^2 \right)
\ee
In order to obtain the free energy, $G$, this has to be integrated over the surface area of the lipid membrane. At constant compression modulus, $K_T^A$, and constant potential $\Psi$, the area change $\Delta A$ equilibrates such that
\begin{eqnarray}
\frac{\partial g}{\partial A}&=&K_T^ A\frac{\Delta A}{A_0^2} - \frac{C_m}{A_0}\left(\frac{\partial \Psi_0}{\partial A}\right)_{\Psi,c}\Psi\\
&&-\frac{1}{2 A_0} \left(\frac{\partial C_m}{\partial A}\right)_{\Psi, c} \left( ( \Psi + \Psi_0)^2 - \Psi_0^2 \right)=0\nonumber
\end{eqnarray}
Therefore, 
\be
\Delta A(\Psi)=A_0\left[\frac{C_m\beta_{\Psi, c}}{K_T^A}\Psi +\frac{\alpha_{\Psi, c}}{K_T^A}\left( ( \Psi + \Psi_0)^2 - \Psi_0^2 \right)\right]\;.
\label{invpiezo1}
\ee
Here, the first linear term is due to the area dependence of the membrane polarization, while the second quadratic term originates from the area dependence of the capacitance.

\subsection*{Flexoelectricity}\label{flexoelectricity}

Let us assume that in eq. (\ref{piezo_n0}) $\Psi$ and $\pi$ are constant. Then we find
\be
dq= \left[(\Psi+\Psi_0)\alpha_{\Psi,A} +C_m\beta_{\Psi,A}\right] d c \;.
\label{flexo_n0}
\ee
This is the (direct) 'flexoelectric effect'. It describes the charging of a capacitor by curvature. If we further assume that the capacitance does not depend on curvature and that the coefficient $\beta_{\Psi,A}$ is constant, we obtain
\be
q (c)= C_m \left(\Psi+\Psi_0 (0)\right)+ C_m\beta_{\Psi,A} c \;,
\label{flexo_n1}
\ee
where $C_m \left(\Psi+\Psi_0 (0)\right)$ is the membrane charge at $c=0$. If the applied potential, $\Psi$, is zero and the polarization in the absence of curvature is also assumed being zero, we obtain
\be
q(c)=C_m\beta_{\Psi,A} c \qquad \mbox{or} \qquad \Psi_0 (c)=\beta_{\Psi,A} c \;.
\label{flexo_n2}
\ee 
Thus, the offset potential $\Psi_0$ is proportional to the curvature. 
This relation is a special case of the flexoelectric effect described in eq.\;(\ref{flexo_n0}). It was previously discussed by Petrov \cite{Petrov1999}. He introduced a flexoelectric coefficient, $f$, which is given by $f\equiv \varepsilon\cdot \beta_{\Psi,A}$. Petrov found experimentally that $f=10^{-18}$\;[C], or $\beta_{\Psi,A}=2.82 \cdot 10^{-8}$ [m] for $\varepsilon = 4\varepsilon_0$, respectively.\\

\noindent\textbf{Inverse flexoelectric effect:} In the absence of a spontaneous curvature, the elastic free energy density of bending is given by $g=\frac{1}{2}K_B c^2$, where $K_B$ is the bending modulus. In the presence of an applied potential  and assuming that $C_m$ does not depend on curvature, the free energy density is given by
\be
g=\frac{1}{2}K_B c^2- \frac{1}{2} \frac{C_m}{A} \left( ( \Psi + \Psi_0)^2 - \Psi_0^2 \right)
\ee
In order to obtain the free energy, $G$, this has to be integrated over the surface area of the lipid membrane. At constant potential $\Psi$, the curvature $c$ equilibrates such that
\be
\frac{\partial g}{\partial c}=K_B c - \frac{C_m}{A}\left(\frac{\partial \Psi_0}{\partial c}\right)_{\Psi,A}\Psi=K_B c - \frac{C_m}{A}\beta_{\Psi,A} \Psi=0
\ee
Therefore, 
\be
c(\Psi)=\frac{C_m}{A}\frac{\beta_{\Psi, A}}{K_B}\Psi=\frac{\varepsilon}{d}\frac{\beta_{\Psi, A}}{K_B}\Psi
\label{invflex1}
\ee
This effect is called the 'inverse flexoelectric effect'. It describes how curvature is induced by an applied potential. It depends on the bending modulus. In melting transitions, the curvature-induction by voltage is enhanced because $K_B$ approaches a minimum \cite{Heimburg1998}. This implies that in the presence of an applied field, the curvature of a membrane changes upon changing the temperature - in particular close to transitions.

Both, the investigation of flexoelectric and inverse flexoelectric effects have been pioneered by Petrov \cite{Petrov1999}. In Petrov's nomenclature, eq. (\ref{invflex1}) assumes the form $c(\Psi)=(f/d \cdot K_B) \Psi$.

\section*{Discussion}
In this publication, we have provided a general thermodynamic treatment of polarization effects on the properties of lipid membranes. When applied to a membrane in an electrolyte, these electric effects can all be related to the charging (or discharging) of capacitors by either potential, curvature or area (or lateral pressure) changes. The latter two effects can lead to an offset potential or a spontaneous polarization. This is important because biological membranes are known to be polar and changes in voltage are generally considered to be central to the understanding of the functioning of cells. We show that a permanent or spontaneous polarization of a membrane influences the properties of a membrane capacitor such that it is discharged at a voltage different from zero. We relate this voltage to an "offset potential". The existence of this potential has the consequence that membrane properties even of chemically symmetric membranes are controlled differently for positive and negative voltages. We derived equations for the piezoelectric and inverse piezoelectric effect. The first considers the change in the offset potential when changing the membrane area. The second considers the change in membrane area by an applied field, which depends on the elastic modulus of the membrane. Finally, we derived general relations for the flexoelectric and the inverse flexoelectric effect. The flexoelectric effect is the change in the offset potential by changing curvature. The inverse flexoelectric effect is the change in curvature induced by an applied potential. We showed that in some simple limiting cases, our derivations lead to relations identical to those of Petrov \cite{Petrov1999}. Petrov pioneered the field of membrane flexoelectricity (e.g., \cite{Petrov1975, Petrov1984, Petrov1986, Petrov1989, Petrov1994, Petrov1999, Petrov2001, Petrov2002a}).

An electric field applied across a lipid membrane generates a force normal to the membrane surface due to the charging of the membrane capacitor. The resulting reduction in membrane thickness is called electrostriction \cite{Heimburg2012}. For fixed membrane dimensions, the electrostrictive force is a quadratic function of voltage. Due to membrane thinning induced by the forces, one finds an increase in membrane capacitance. This has been demonstrated for symmetric black lipid membranes made from phosphatidylethanolamines (Fig. \ref{Figure1d}, \cite{Alvarez1978}. However, for an asymmetric membrane made of charged lipids on one side and zwitterionic lipids on the other side (thus displaying polarity) the minimum capacitance is found at a voltage different from zero (Fig. \ref{Figure1d}, \cite{Alvarez1978}). This indicates that a permanent electric polarization of the membrane influences the capacitive properties of a membrane. This has also been found in biological preparations. Human embryonic kidney cells display an offset potential of $-51$ mV \cite{Farrell2006}. This indicates that the capacitance in electrophysiological models such as the Hodgkin-Huxley model \cite{Hodgkin1952b} is incorrectly used because offset potentials are not considered. However, it is very likely that the offset potentials are closely related to the resting potentials of membranes. It should also be noted that the capacitance is typically dependent on the voltage. This effect has also not been considered in classical electrophysiology models. We treat that here in terms of a 'capacitive susceptibility' (eq.\;(\ref{cap_susc1}), cf. \cite{Heimburg2012}).

Electrostrictive forces also influence melting transitions of lipid membranes. Since the fluid state of the membrane displays a smaller thickness than the gel phase, an electrostrictive force will shift the state of the membrane towards the fluid state. Heimburg \citep{Heimburg2012} calculated a decrease of the melting temperature, $T_m$, which is a quadratic function of voltage. Since the membrane was considered being symmetric, the largest $T_m$ is found at $\Psi=0$. Here, we showed that a membrane which displays a spontaneous polarization in the absence of an applied electric field possesses an offset potential, $\Psi_0$, in the free energy (\eref{el-striction}). The respective equation contains the term $((\Psi+\Psi_0)^2-\Psi_0^2)=\Psi^2+2\Psi \Psi_0$, which is approximately linear for $\Psi\ll\Psi_0$ (eq. (\ref{G_el_1}). In fact,  Antonov and collaborators found a linear dependence of the melting temperature on voltage \citep{Antonov1990}. 
This indicates that the membranes studied by Antonov and collaborators \cite{Antonov1990} were polar. 

Antonov's experiment determined the voltage-dependence of the melting temperature by measuring the permeability chan{-}ges in the transition. It is well known that membranes display maximum conductance in lipid phase transitions \cite{Papahadjopoulos1973, Blicher2009}. Furthermore, it has been found that membranes can form pores that appear as quantized conduction event upon the application of potential difference across the membrane \cite{Antonov1980, Kaufmann1983b, Blicher2009, Heimburg2010, Mosgaard2013b}. The likelihood to form a pore is thought to be proportional to the square of the applied electric potential \cite{Winterhalter1987, Glaser1988}. This is based on the assumption that an increase in voltage thins the membrane and eventually leads to an electric breakdown linked to pore formation. Laub et al. \cite{Laub2012} found that the current-voltage (I-V) relation for a chemically symmetric phosphatidylcholine membrane patch formed on the tip of a glass pipette was a non-linear function of voltage which was not symmetric around $\Psi=0$, but rather outward rectified.
Blicher et al. \cite{Blicher2013} proposed that a voltage offset can explain the outward-rectification. They proposed that the free energy difference between an open and a closed pore, $\Delta G_p$,  can be expressed by
\be
\Delta G_p=\Delta G_{p,0}+\alpha (V-V_0)^2 \;,
\ee
where $\Delta G_{p,0}$ and $\alpha$ are coefficients and $V_0$ is a voltage offset. This equation has the same analytic form as used here for the electrostatic free energy ($G=-(C_m/2)((\Psi+\Psi_0)^2-\Psi_0^2)$). Assuming that the equilibrium constant between an open and a closed form of a membrane pore is given by $K_p=$ \linebreak$\exp(-\Delta G_p/kT)$ and the likelihood of finding an open pore is given by $P_{open}=K_p/(1+K_p)$, Blicher and collaborators concluded that the I-V relation could be expressed as
\be
I=\gamma_p P_{open}V
\ee
This relation perfectly fitted the experimental current-voltage data. Thus, inward and outward rectified I-V profiles can be found in pure lipid membranes in the complete absence of proteins. They find their origin in the polarization of the membrane.\\

Here, we investigated two possible mechanisms that can give rise to spontaneous polarization in the absence of an applied field, which both break the symmetry of the membrane. The first (flexoelectricity) acts by allowing the membrane to be curved (thus introducing a curvature, $c$) and a difference of the lateral tension within the two monolayers. The second mechanism acts by assuming a chemically or physically asymmetric lipid composition on the two leaflets.  An example for a physically asymmetric membrane is a situation where one monolayer is in a fluid state while the other monolayer is in a gel state. Chemical asymmetry assumes a different lipid composition on the two sides of the membrane. The magnitude of the resulting offset, $\Psi_0$, is strongly influenced by experimental conditions such as the lipid composition, salt concentration, pH, or the presence of divalent ions. Permanent polarization of the lipids can not only lead to an electrical offset but also to an enhanced dielectric constant. For biological membranes, polarization asymmetries can originate from any constituting element of the membrane including integral membrane proteins. We can also speculate that other membrane adhesive molecules with large dipoles can be used to create an asymmetric membrane, e.g., soluble proteins or lipid-associated molecules such as long-chain sugars. Depending on the nature of the asymmetry, the system can display piezoelectric properties.

The offset potential can have interesting consequences for capacitive currents. The charge on a capacitor is given by $q=C_m(\Psi+\Psi_0)$. Therefore, for constant $\Psi_0$ the capacitive current is given by
\be
I_c (t)=\frac{dq}{dt}=C_m\frac{d\Psi}{dt}+\left(\Psi+\Psi_0\right)\frac{dC_m}{dt}
\label{Vclamp}
\ee
For a positive change in potential, the first term in eq. (\ref{Vclamp}) is positive and leads to a positive current. If the change in voltage happens instantaneously, the corresponding current peak is very short. The second term describes the temporal change in capacitance induced by the voltage change. It depends on the relaxation time of the membrane capacitance, which close to transitions can range from milliseconds to seconds. Thus, it can be distinguished from the first term. Let us consider the situation shown in Fig. \ref{Figure3} ($\Psi_0=70$\;mV, T=314.5 K) with a membrane capacitance of $\approx$ 1 \textmu F/cm$^2$. Here, a jump from $\Psi=-70$ mV to $\Psi=-10$ mV yields a positive change in capacitance of $\Delta C_m=2.6$ nF/cm$^2$. If the offset potential were $\Psi_0=-70$ mV instead, the same jump would change the capacitance by $\Delta C_m =-7.8$ nF/cm$^2$. Therefore, the second term in eq. (\ref{Vclamp}) is positive in the first situation but negative in the second situation. For this reason, depending on the offset potential and holding potential, the capacitive current associated to the second term in eq. (\ref{Vclamp}) can go along the applied field or against the applied field. Similarly, for a jump in potential of +60 mV, the capacitive current would depend on the holding potential before the jump. For $\Psi_0 = 70$ mV, the change in capacitance is $\Delta C_m=-2.6$ nF/cm$^2$ for a jump from $-130$ mV to $-70$ mV. It is $\Delta C_m=+8.9$ nF/cm$^2$ for a jump from $+70$ mV to $+130$ mV. The typical time-scale of processes in biomembranes is a few milliseconds to a few ten milliseconds.  It can be different for different voltages. Thus, slow currents on this time-scale against an applied field can originate from voltage-induced changes in lipid membrane capacitance. If the offset-potential also depends on voltage, this situation is more complicated.

Flexoelectric and piezoelectric phenomena have also be considered to be at the origin of an electromechanical mechanism for nerve pulse propagation \cite{Heimburg2005c}. In 2005, Heimburg and Jackson proposed that the action potential in nerves consists of an electromechanical soliton. The nerve pulse is considered as a propagating local compression of the membrane with a larger area density. According to the piezoelectric effect treated here (eq.\;(\ref{piezo_n1}), a change in membrane area can lead to the charging of the membrane capacitor. Alternatively, due to the inverse piezoelectric effect a change in the applied membrane potential can induce area changes (eq.\;(\ref{invpiezo1}) and thus induce a density pulse. The inverse piezoelectric effect is very dependent on the lateral compressibility of a membrane. Thus, is is largely enhanced in the melting transition where the compressibility is high. Further, these effects will largely depend on membrane polarization.

Finally, it should be mentioned that some of the polarization effects on artificial membranes are not very pronounced because changes in polarization due to changes in area are not very large. For instance, a voltage change of 200\;mV changes the transition temperature by only 0.12 K. However, the absolute magnitude of the effect largely depends on offset polarizations. These could be influenced by lipid-membrane-associated molecules (such as proteins) with large dipole moments.


\section*{Conclusion}
Here, we provided a unified thermodynamic framework for capacitive changes, piezoelectricity and flexoelectricity. It treats all of these effects in terms of the electric field, $E$, and the electric displacement, $D$. We show that a spontaneous membrane polarization leads to offset potentials that form the origin for a number of interesting membrane phenomena, including voltage-dependent changes in capacitance, voltage-induced curvature, rectified current-voltage relations for membrane conductance, and capacitive currents against the applied field.\\


\textbf{Author Contributions:}
LDM, KAZ and TH developed the theory, made the calculations, designed and wrote the article together.\\


\textbf{Acknowledgments:} We thank to Prof. Andrew D. Jackson from the Niels Bohr International Academy for useful discussions and for a critical reading of the manuscript. This work was supported by the Villum foundation (VKR 022130).


\footnotesize{

}
\end{document}